# Continuously Ordered Hierarchies of Algorithmic Information in Digital Twinning and Signal Processing

Yannik N. Böck*, Holger Boche*†, and Frank H.P. Fitzek‡§

*Abstract*— We consider a fractional-calculus example of a continuous hierarchy of algorithmic information in the context of its potential applications in digital twinning. *Digital twinning* refers to different emerging methodologies in control engineering that involve the creation of a digital replica of some physical entity. From the perspective of *computability theory*, the problem of ensuring the digital twin's *integrity* – i.e., keeping it in a state where it matches its physical counterpart – entails a notion of *algorithmic information* that determines which of the physical system's properties we can reliably deduce by algorithmically analyzing its digital twin. The present work investigates the *fractional calculus* of *periodic functions* – particularly, we consider the *Wiener algebra* – as an exemplary application of the algorithmic-information concept. We establish a continuously ordered hierarchy of algorithmic information among spaces of periodic functions – depending on their fractional degree of smoothness – in which the ordering relation determines whether a certain representation of some function contains "more" or "less" information than another. Additionally, we establish an analogous hierarchy among $\ell^p$-*spaces*, which form a cornerstone of (traditional) digital signal processing. Notably, both hierarchies are (mathematically) "dual" to each other. From a practical perspective, our approach ultimately falls into the category of formal verification and (general) *formal methods*.

## I. INTRODUCTION

*Digital twinning* refers to a class of emerging methodologies in different areas of engineering. Methods that belong to this class involve a *physical entity* to be twinned, a *physical process* that governs the entity's time evolution, and a digital replica of the entity, called the *digital twin* [1], [2]. Often, the physical entity is *abstract*, such as a *network* of moving agents (see e.g. [3]). The *integrity* of the joint system of physical entity and digital twin corresponds to the "conformity" of both parts in their (regarding the respective technological application) *relevant* properties.

Near-future *cyber-physical systems* – especially in the context of Industry 4.0 and the upcoming 6G communications standard – will involve digital twinning as a key enabler of novel *autonomous-control* technologies. Any digital twin ultimately serves as the subject of an automated "virtual analysis" to predict its physical counterpart's evolution and implement elaborate control and decision-making strategies [1]. From a mere conceptual point of view, digital twinning thus strongly overlaps with the *internal-model principle* [4]. However, to the best of the authors' knowledge, formal theories of digital twinning are far less established than formal theories of the internal-model principle. At this stage, digital-twinning methods appear to be characterized primarily by informal qualities rather than a mathematical framework.

The sound operation of a digital-twinning system requires the digital twin to match the physical entity's relevant properties, i.e., the joint system must maintain its integrity. As near-future cyber-physical systems will have significant potential of affecting sensitive aspects of human well-being [5], ensuring the integrity of such systems per design-basis – which we will henceforth refer to as the *integrity problem* – is necessary. In our previous work, we have demonstrated the integrity problem's solvability to depend crucially on *how* the underlying software *represents* the replicated physical entity [6]. The representation method's formal specification generally depends on the physical process's mathematical characterization and is thus highly application-specific.

From the perspective of *computability theory*, each formally specified representation method defines a certain *type* of *algorithmic information*, which then determines whether we can solve the integrity problem for a given decision-making or control task. We analyze an *effective* (i.e., computable) *fractional calculus* for the *(real) Wiener algebra* of *periodic functions* $f : [0, 2\pi) \to \mathbb{R}$ in this regard. Periodic functions and fractional calculus are vital to the mathematical modeling of physical processes. We also apply our framework to the $\ell^p$-spaces of *asymptotically vanishing sequences*, which form a cornerstone of (traditional) digital signal processing. We argue that our framework falls conceptually into the category of *formal methods*. This may not be immediately

The work of *Yannik N. Böck* and *Holger Boche* was supported in part by German Federal Ministry of Education and Research (BMBF) within the National Initiative on 6G Communication Systems through the Research Hub 6G-life under Grant 16KISK002. The work of *Frank H.P. Fitzek* was supported in part by German Research Foundation (DFG, Deutsche Forschungsgemeinschaft) as part of Germany's Excellence Strategy – EXC2050/1 – Project ID 390696704 – Cluster of Excellence "Centre for Tactile Internet with Human-in-the-Loop" (CeTI) of Technische Universität Dresden. *The authors* were further supported in part by the BMBF in the program "Souverän. Digital. Vernetzt." – Joint project 6G-life, project identification number 16KISK001K

* Technical University of Munich, School of Computation, Information and Technology, Department of Computer Engineering, Chair of Theoretical Information Technology, D-80333 Munich, Germany {yannik.boeck, boche}@tum.de

† BMBF Research Hub 6G-life Munich, D-80333; Munich Center for Quantum Science and Technology (MCQST), D-80799; Munich Quantum Valley (MQV) D-80807

‡ Telekom Chair of Communication Networks, Technische Universität Dresden (TUD), D-01062 Dresden, Germany frank.fitzek@tu-dresden.de

§ BMBF Research Hub 6G-life Dresden, D-01062; TUD Cluster of Excellence "Centre for Tactile Internet with Human-in-the-Loop (CeTI)," D-01062; TUD 5G Lab Germany, D-01062

For timeliness, we restrict ourselves to providing proof sketches for Theorems 1 and 2; the methods for proving Theorems 3 and 4 are analogous, respectively. The article will be updated to include complete proofs of all presented results as soon as possible.

evident, as the relevant literature employs terminology and notation vastly different from ours. We will thus comment on the reasoning behind this claim in Sections II and III.

*A. Notation, Nomenclature and General Remarks*

For generic sets $\mathcal{A}$ and $\mathcal{B}$, a *partial function* $f : \mathcal{A} \supseteq\to \mathcal{B}$ is of the form $f : D(f) \to \mathcal{B}$, $D(f) \subseteq \mathcal{A}$. We call $D(f)$ the *domain* of $f$. If $D(f) = \mathcal{A}$, we call $f$ a *total* function. Since the inclusion $D(f) \subseteq \mathcal{A}$ includes the *improper* case, every total function is also partial, but not vice versa.

Throughout this paper, we consider functions $f : \mathcal{I} \to \mathbb{R}$, where $\mathcal{I} := [0, 2\pi) \subset \mathbb{R}$. For $m \in \mathbb{N}$, we define

$$\sin_m : \mathcal{I} \to \mathbb{R},\ t \mapsto \sin_m(t) := \sin(mt),$$
$$\cos_m : \mathcal{I} \to \mathbb{R},\ t \mapsto \cos_m(t) := \cos(mt).$$

Further, we use the notation $\exp_m(\omega) := m^\omega$, where $m \in \mathbb{N}$, $\omega \in \mathbb{Q}$. Note that $\exp_0(0)$ is *undefined*.

We consider different *computable* objects and sequences, each of which belongs to some countable set $\Xi_\mu$. In this context, 'computable in $\Xi_\mu$' and '$\Xi_\mu$-computable' are synonymous terminology. Note that $\Xi_\mu$ always denotes a countable subset of some *generic* set $\Xi$. Other than being well-defined, we do *not* impose any restrictions on $\Xi_\mu$ and $\Xi$.

If an expression $\text{expr}(\cdot)$ of some form defines the elements of a sequence $(a_n)_{n \in \mathbb{N}}$, we employ the notation

$$(a_n)_{n \in \mathbb{N}} : a_n = \text{expr}(n)$$

for the sequence's definition. For example, for $\text{expr}(\cdot) \equiv \sqrt{\cdot}$, we may write $(a_n)_{n \in \mathbb{N}} : a_n = \sqrt{n}$.

## II. MOTIVATION: A FORMAL THEORY OF INTEGRITY

Subsequently, we discuss a concise operationalization of algorithmic information in the digital-twinning and integrity context. Henceforth, we will not explicitly distinguish between a physical entity an the physical process that governs the entity's time-evolution, as it is not necessary with regards to our theory's purely mathematical aspects. Formally, a physical process is a sequence $(v_t)_{t \in \mathbb{R}} \subset \mathbb{B}$, where $\mathbb{B}$ is a Banach space with norm $\|\cdot\|_\mathbb{B}$, s.t.

$$\sup_{t \in \mathbb{R}} \|v_t\|_\mathbb{B} < \infty \quad \text{and} \quad \forall t \in \mathbb{R}.\ \lim_{s \to t} \|v_t - v_s\|_\mathbb{B} = 0.$$

With $\|(v_t)_{t \in \mathbb{R}}\|_\infty := \sup_{t \in \mathbb{R}} \|v_t\|_\mathbb{B}$, the set of such sequences is a Banach space itself, which we denote by $\mathcal{C}^0(\mathbb{R}, \mathbb{B})$. Often, $(v_t)_{t \in \mathbb{R}}$ satisfies an equation of the form

$$\mathfrak{D}[(v_t)_{t \in \mathbb{R}}] = w \in \mathbb{W} \quad \text{with} \quad \mathfrak{D} : \mathcal{C}^0(\mathbb{R}, \mathbb{B}) \supseteq\to \mathbb{W},$$

where $\mathfrak{D}$ is a linear operator and $\mathbb{W}$ is another Banach space that may or may not be similar to $\mathcal{C}^0(\mathbb{R}, \mathbb{B})$ in structure. As indicated in Section I, the time evolution of a network of moving agents (see e.g. [3]) is an example of a physical process.

Since digital computing is fundamentally discrete, any digital computer can distinguish at most countably many different "representations" of physical entities. Thus, any method of representing physical processes on digital hardware requires choosing a countable subset $\mathbb{B}_\mu \subset \mathbb{B}$, the elements of which we represent by means of a surjective mapping $\mathsf{N}_\mathbb{B} : \mathbb{N} \supseteq\to \mathbb{B}_\mu$. We call such a mapping a *numbering* (c.f. Section III for the formal context). At the time being, we are primarily interested in the set $\mathbb{B}_\mu$. The numbering $\mathsf{N}_\mathbb{B}$ will become relevant further below. Given a physical process $(v_t)_{t \in \mathbb{R}}$, its digital twin consists of a discrete-time sequence $(\hat{v}_n)_{n \in \mathbb{N}} \subset \mathbb{B}_\mu$, where $\hat{v}_n$ corresponds to $v_{nT}$ for all $n \in \mathbb{N}$ and some cycle time $T \in \mathbb{R}$, $T > 0$.

As indicated in Section I, we want to employ the digital twin $(\hat{v}_n)_{n \in \mathbb{N}}$ to algorithmically predict properties of the physical entity $(v_t)_{t \in \mathbb{R}}$ in most practical applications (c.f. [1], [5]). Formally, any such property corresponds to some binary relation $\rightleftharpoons_\mathbb{B} \subseteq \mathbb{B}_\mu \times \mathbb{B}$. Note that $\rightleftharpoons_\mathbb{B}$ is strictly speaking a property of the *joint system* consisting of physical entity *and* digital twin. Since the prediction algorithm will not have access to the physical entity, it is generally necessary to allow $\rightleftharpoons_\mathbb{B}$ to depend on both parts of the joint system. Consider, for example, a network of $M \in \mathbb{N}$ point-like mobile agents moving in the two-dimensional Euclidean plane, i.e.,

$$\mathbb{B} \equiv (\mathbb{R}^2)^{\times M}, \quad v_t \equiv (\vec{x}_{t,1}, \ldots, \vec{x}_{t,M}), \quad t \in \mathbb{R}.$$

For *collision avoidance*, it may be necessary to ensure that no two agents fall short of a minimum distance $r_{\min} > 0$ from each other [3]. In this case, we may consider

$$\hat{v}_n \rightleftharpoons_\mathbb{B} v_t := \big((\|\hat{v}_n - v_t\|_\mathbb{B} \leq \epsilon) \Rightarrow \ldots$$
$$\ldots (r_{\min} < \min_{1 \leq m < l \leq M} \|\vec{x}_{t,m} - \vec{x}_{t,l}\|_\mathbb{B})\big). \quad (1)$$

Note that if $\|\hat{v}_n - v_t\|_\mathbb{B} \geq \epsilon$, the *antecedent* in (1) is *false*. Thus, $\hat{v}_n \rightleftharpoons_\mathbb{B} v_t$ is necessarily *true* in this case.

In the context of our theory, a prediction algorithm is a *partial* mapping that is *zero* everywhere on its domain. That is, prediction algorithms are of the form

$$\texttt{Alg} : \mathbb{B}_\mu \supseteq\to \{0\} \quad \text{with} \quad D(\texttt{Alg}) \subseteq \mathbb{B}_\mu. \quad (2)$$

Note that $D(\texttt{Alg})$ uniquely determines $\texttt{Alg} : \mathbb{B}_\mu \supseteq\to \{0\}$. Further, $D(\texttt{Alg})$ and exhibits a special interpretation in terms of the sequential nature of classical computations: It consists of those inputs that lead the stepwise procedure underlying $\texttt{Alg}$ to eventually *terminate*. The "value" $\texttt{Alg}(\hat{v})$ is well-defined (and equal to zero) only if $\hat{v}$ is an element of $D(\texttt{Alg})$. Precisely, the halting of $\texttt{Alg}$ for input $\hat{v}$ serves as either a *necessary* or *sufficient* condition for $\hat{v}_n \rightleftharpoons_\mathbb{B} v_t$ to hold.

> **GO Integrity** (of $\texttt{Alg} : \mathbb{B}_\mu \supseteq\to \{0\}$ w. resp. to $\rightleftharpoons_\mathbb{B} \subseteq \mathbb{B}_\mu \times \mathbb{B}$).
> $\forall \hat{v} \in \mathbb{B}_\mu, v \in \mathbb{B}.\ ((\texttt{Alg}\text{ halts for input }\hat{v}) \Rightarrow (\hat{v} \rightleftharpoons_\mathbb{B} v))$
> **NO-GO Integrity** (of $\texttt{Alg} : \mathbb{B}_\mu \supseteq\to \{0\}$ w. resp. to $\rightleftharpoons_\mathbb{B} \subseteq \mathbb{B}_\mu \times \mathbb{B}$).
> $\forall \hat{v} \in \mathbb{B}_\mu, v \in \mathbb{B}.\ ((\hat{v} \rightleftharpoons_\mathbb{B} v) \Rightarrow (\texttt{Alg}\text{ halts for input }\hat{v}))$

The terminology refers to the GO/NO-GO analysis of safety-critical systems:

- If $\rightleftharpoons_\mathbb{B}$ is a condition that *must* hold for the system to operate safely, we allow operation only if $\texttt{Alg}$ has verified that $\rightleftharpoons_\mathbb{B}$ is fulfilled;
- If $\rightleftharpoons_\mathbb{B}$ is a condition that *must not* hold for the system to operate safely, we require that $\texttt{Alg}$ will certainly indicate whenever $\rightleftharpoons_\mathbb{B}$ is fulfilled.

Observe that we can always *trivially* maintain *GO* and *NO-GO* integrity by choosing $\texttt{Alg}$ to *never* or *always* terminate, respectively, which corresponds to not operating the system at all. On the other hand, consider $\texttt{GO}_{\rightleftharpoons}, \texttt{NO}_{\rightleftharpoons} : \mathbb{B}_\mu \supseteq\to \{0\}$ defined through

$$D(\texttt{GO}_{\rightleftharpoons}) := \{\hat{v} \in \mathbb{B}_\mu | \forall v \in \mathbb{B}.\ \hat{v} \rightleftharpoons_\mathbb{B} v\},$$
$$D(\texttt{NO}_{\rightleftharpoons}) := \{\hat{v} \in \mathbb{B}_\mu | \exists v \in \mathbb{B}.\ \hat{v} \rightleftharpoons_\mathbb{B} v\}.$$

In order for $\texttt{Alg}$ to maintain GO or NO-GO integrity, we must have $D(\texttt{Alg}) \subseteq D(\texttt{GO}_{\rightleftharpoons})$ or $D(\texttt{NO}_{\rightleftharpoons}) \subseteq D(\texttt{Alg})$, respectively. Thus, $D(\texttt{GO}_{\rightleftharpoons})$ and $D(\texttt{NO}_{\rightleftharpoons})$ characterize the "least trivial" extent to which we can maintain integrity.

So far, we have evaded specifying the term "algorithm", except for the minimum requirement (2). At some point, however, we must implement $\texttt{Alg}$ in one way or another on digital hardware. As indicated above, digital computing ultimately operates on the natural numbers – or, from a practical perspective, bit-string representations thereof. In Section III, we will introduce $\mu$-*recursive functions*, which form a special class of partial functions of the form $g : \mathbb{N} \times \cdots \times \mathbb{N} \supseteq\to \mathbb{N}$. Particularly, they precisely form the class of functions we can compute by means of digital hardware. In contrast, $\mathbb{B}$, $(v_n)_{n \in \mathbb{N}}$, and $\rightleftharpoons_\mathbb{B}$ are abstract mathematical concepts. In order to make these concepts the subject of digital computing, we must employ a suitable numbering. Formally, an implementation of some algorithm $\texttt{Alg} : \mathbb{B}_\mu \supseteq\to \{0\}$ with respect to some numbering $\mathsf{N}_\mathbb{B} : \mathbb{N} \supseteq\to \mathbb{B}_\mu$ is a $\mu$-recursive function $g : \mathbb{N} \supseteq\to \{0\}$ that satisfies

$$D(g) = \{n \in D(\mathsf{N}_\mathbb{B}) \mid \mathsf{N}_\mathbb{B}(n) \in D(\texttt{Alg})\}.$$

Thus, the computation underlying $g$ must terminate for input $n \in \mathbb{N}$ if and only if $\mathsf{N}_\mathbb{B}(n) = \hat{v}$ for some $\hat{v} \in D(\texttt{Alg})$.

Assume we have $\mathsf{N}_\mathbb{B}(n) = \hat{v}$ for some $n \in \mathbb{N}, \hat{v} \in \mathbb{B}_\mu$. In the terminology of recursion theory and computable analysis, we call $n$ a *realizer* of $\hat{v}$. Informally, $n$ "represents" the abstract object $\hat{v} \in \mathbb{B}_\mu$ with respect to the (formally specified) "representation method" $\mathsf{N}_\mathbb{B}$. This paper's focal question then presents itself as follows.

**Algorithmic Information.** *Given a formally specified method of digitally representing the objects of some (abstract) set, which of an object's properties can we reliably deduce by algorithmically analyzing any of the object's representations?*

In the context of digital twinning, we interpret reliable deduction in terms of GO or NO-GO integrity: We can "reliably deduce" a property $\rightleftharpoons_\mathbb{B} \subseteq \mathbb{B}_\mu \times \mathbb{B}$ if, respectively, the corresponding algorithm $\texttt{GO}_{\rightleftharpoons}$ or $\texttt{NO}_{\rightleftharpoons}$ exhibits an implementation with respect to the chosen numbering $\mathsf{N}_\mathbb{B}$. On the other hand, we may naturally interpret the class of algorithms $\texttt{Alg} : \mathbb{B}_\mu \supseteq\to \{0\}$ that exhibit an implementation with respect to $\mathsf{N}_\mathbb{B}$ as the "information" contained in the elements of $D(\mathsf{N}_\mathbb{B})$. At least qualitatively, we can formalize this concept by a *quasiorder* on the set of numberings of $\mathbb{B}_\mu$. For $\mathsf{N}_\mathbb{B}, \mathsf{N}'_\mathbb{B} : \mathbb{N} \supseteq\to \mathbb{B}_\mu$, we write

- $\mathsf{N}_\mathbb{B} \preceq_\mu \mathsf{N}'_\mathbb{B}$ if every prediction algorithm $\texttt{Alg} : \mathbb{B}_\mu \supseteq\to \{0\}$ that exhibits an implementation with respect to $\mathsf{N}_\mathbb{B}$ also exhibits an implementation with respect to $\mathsf{N}'_\mathbb{B}$;
- $\mathsf{N}_\mathbb{B} \simeq_\mu \mathsf{N}'_\mathbb{B}$ if $\mathsf{N}_\mathbb{B} \preceq_\mu \mathsf{N}'_\mathbb{B}$ *and* $\mathsf{N}'_\mathbb{B} \preceq_\mu \mathsf{N}_\mathbb{B}$;
- $\mathsf{N}_\mathbb{B} \prec_\mu \mathsf{N}'_\mathbb{B}$ if $\mathsf{N}_\mathbb{B} \preceq_\mu \mathsf{N}'_\mathbb{B}$ but *not* $\mathsf{N}'_\mathbb{B} \preceq_\mu \mathsf{N}_\mathbb{B}$;
- $\mathsf{N}_\mathbb{B} \not\asymp_\mu \mathsf{N}'_\mathbb{B}$ if *neither* $\mathsf{N}_\mathbb{B} \preceq_\mu \mathsf{N}'_\mathbb{B}$ *nor* $\mathsf{N}'_\mathbb{B} \preceq_\mu \mathsf{N}_\mathbb{B}$.

If we have $\hat{v} = \mathsf{N}_\mathbb{B}(n) = \mathsf{N}'_\mathbb{B}(m)$ for $\hat{v} \in \mathbb{B}_\mu, n, m \in \mathbb{N}$, and numberings $\mathsf{N}_\mathbb{B}, \mathsf{N}'_\mathbb{B} : \mathbb{N} \supseteq\to \mathbb{B}_\mu$ that satisfy $\mathsf{N}_\mathbb{B} \preceq_\mu \mathsf{N}'_\mathbb{B}$, we say that *"n contains at most as much information about $\hat{v}$ as m does."* Interpretations of the remaining possible combinations are straightforward.

The subsequent Section III will put our framework on mathematically rigorous grounds through the theory of $\mu$-recursive functions and *computable analysis*. As indicated above, the community of computer science widely considers $\mu$-recursive functions as the definitive mathematical characterization of digital computing. Nevertheless, applied numerics much more commonly (and usually implicitly) considers the *real RAM* (RRAM) model [7, Section 1.4, p. 26ff], which features *infinitely precise* real-valued arithmetic operations. In many cases, this makes the elements of $\mathbb{B}$ directly accessible to the RRAM machine, obsoleting the set $\mathbb{B}_\mu$.

Albeit sufficient for many applications, the finite precision of real-world floating-point hardware places digital implementations of RRAM algorithms on a heuristic foundation. Let $w_0, \hat{w}_1 \in \mathbb{B}$ be arbitrary but fixed and assume we can prove

$$\forall \hat{v} \in \mathbb{B}.\ \big((\texttt{Alg}_\mathrm{R} \text{ halts for input } \hat{v}) \Rightarrow (\hat{v} \rightleftharpoons_\mathbb{B} w_0)\big) \quad (3)$$

for some RRAM algorithm $\texttt{Alg}_\mathrm{R} : \mathbb{B} \supseteq\to \{0\}$. Furthermore, assume upon implementation, we observe $\texttt{Alg}_\mathrm{R}$ to halt for input $\hat{w}_1$. Even under the premise of correctly operating hardware and despite our success in proving (3), it is *infeasible* to conclude $\hat{w}_1 \rightleftharpoons_\mathbb{B} w_0$ in the RRAM setting. This is *not* the case for the computable-analysis approach. If we can prove

$$\forall \hat{v} \in \mathbb{B}_\mu.\ \big((\texttt{Alg}_\mathrm{T} \text{ halts for input } \hat{v}) \Rightarrow (\hat{v} \rightleftharpoons_\mathbb{B} w_0)\big)$$

for some Turing algorithm $\texttt{Alg}_\mathrm{T} : \mathbb{B} \supseteq\to \{0\}$ and – assuming $\hat{w}_1 \in \mathbb{B}_\mu$ and correctly operating hardware – observe any of its possible implementations to halt for input $\hat{w}_1$, we can in principle extract a mathematical proof of $\hat{w}_1 \rightleftharpoons_\mathbb{B} w_0$ from the computation the algorithm performs upon receiving its input. The computable-analysis approach thus entails formal guarantees for system behavior, which places our framework into the category of *formal methods*.

### III. BACKGROUND: COMPUTABLE ANALYSIS

Subsequently, we provide a brief overview of *recursion theory* and *computable analysis*, both of which are instrumental to the further parts of this work. For comprehensive treatments, we refer to [8]–[12].

The set of $\mu$-*recursive functions* [13] is the smallest set of partial functions $g : \mathbb{N} \times \cdots \times \mathbb{N} \supseteq\to \mathbb{N}$ that

- contains the *successor function*, all *constant functions*, and all *projection functions* and
- is closed with respect to *composition*, *primitive recursion*, and *unbounded search*

(c.f. [8, Definition 2.1, p. 8, Definition 2.2, p. 10] for details). Furthermore, the set of $\mu$-recursive functions coincides with the set of functions we can compute by means of a *universal Turing machine* [14], [15]. (see [16] for the proof of their equivalence). *Turing machines* form a mathematical abstraction of digital computing, i.e., they formalize our intuitive understanding of what we call an *algorithm*. The *Church-Turing thesis* [8, *Author's Preface*], which is widely accepted in the community of computer science, asserts that this formalization is definitive: If, in theory, we *cannot* solve a mathematical problem by means of a Turing machine, we *cannot* solve it on a real-world digital computer.

As a consequence of the *halting problem* [8, Chapter I, Section 4, p. 18ff], the domain $D(g)$ of a recursive function $g : \mathbb{N} \supseteq \to \mathbb{N}$ is *not* generally a *recursive* set. That is, the *indicator function* $\mathbb{1}_{D(g)} : \mathbb{N} \to \{0,1\}$ is *not* necessarily $\mu$-recursive. Particularly, we call a set $\mathcal{A} \subseteq \mathbb{N}$ *recursively enumerable* if we have $\mathcal{A} = D(g)$ for any $\mu$-recursive function $g$. If $\mathcal{A}$ and $\mathbb{N} \setminus \mathcal{A}$ are *both* recursively enumerable, we refer to $\mathcal{A}$ as *recursive*. As indicated above, a set is recursive if and only if its indicator function is $\mu$-recursive.

Consider a sequence $\mathcal{A}_0, \mathcal{A}_1, \mathcal{A}_2, \ldots$ of non-empty recursively enumerable sets. We call such a sequence *computable* if there exists a $\mu$-recursive function $g : \mathbb{N} \times \mathbb{N} \supseteq \to \mathbb{N}$ s.t. for all $n \in \mathbb{N}$, we have $\mathcal{A}_n = \{m \in \mathbb{N} | (n,m) \in D(g)\}$. Note that the following statements are *equivalent*:

- there exists a $\mu$-recursive function $g : \mathbb{N} \times \mathbb{N} \supseteq \to \mathbb{N}$ s.t. for all $n \in \mathbb{N}$, we have $\mathcal{A}_n = \{m \in \mathbb{N} | (n,m) \in D(g)\}$;
- there exists a (total) $\mu$-recursive function $h : \mathbb{N} \times \mathbb{N} \to \mathbb{N}$ s.t. for all $n \in \mathbb{N}$, we have $\mathcal{A}_n = \{h(n,m) | m \in \mathbb{N}\}$.

For a comprehensive analysis of recursively enumerable sets and their properties (such as the equivalences above), we once more refer to [8].

For some countable set $\Xi_\mu$, consider a binary relation $\mathbf{r}_\mu \subseteq \mathbb{N} \times \Xi_\mu$ (we write $n \ \mathbf{r}_\mu \ \xi$ rather than $(n, \xi) \in \mathbf{r}_\mu$ in most cases) s.t. for all $\xi \in \Xi_\mu$, there exists $n \in \mathbb{N}$ s.t. $n \ \mathbf{r}_\mu \ \xi$. We call

$$\Xi_\mu^\star := \{(\xi, \mathbf{r}_\mu) \mid \xi \in \Xi_\mu\}$$

an *assembly* and any $n \in \mathbb{N}$ satisfying $n \ \mathbf{r}_\mu \ \xi$ a *realizer* of $\xi \in \Xi_\mu$. If we have

$$\mathbf{r}_\mu = \{(n, \mathsf{N}_\Xi(n)) \mid n \in D(\mathsf{N}_\Xi)\}$$

for some (surjective) mapping $\mathsf{N}_\Xi : \mathbb{N} \supseteq \to \Xi_\mu$, we call $\Xi_\mu^\star$ a *modest set* and $\mathsf{N}_\Xi$ a *numbering* of $\Xi_\mu$.

Throughout this work, we will not make explicit use of assemblies and modest sets. Rather, we will be concerned with sets $\Xi_\mu$ as above and the relations between different numberings thereof. However, as we have previously (c.f. Sections I and II) stressed the relationship of our theory to formal methods, introducing assemblies and modest sets is necessary for reasons of Completeness. Specifically, assemblies and modest sets form the basis of the *type-theoretic* formalization of computable analysis. This formalization, for its part, establishes the link between algorithms and formal mathematical proofs as indicated in Section II. The interested reader may find a comprehensive exposition in [10].

Consider a generic conutable set $\Xi_\mu$ and numberings $\mathsf{N}_\Xi, \mathsf{N}'_\Xi : \mathbb{N} \supseteq \to \mathbb{N}$. In Section II, we have introduced a quasiorder on the set of numberings associated to $\Xi_\mu$. *Algorithmic reducibility* [11, Section 2.3, p. 33ff] induces another quasiorder on this set: We write $\mathsf{N}_\Xi \leq_\mathrm{T} \mathsf{N}'_\Xi$ whenever there exists a $\mu$-recursive functions $g : \mathbb{N} \supseteq \to \mathbb{N}$ s.t. $D(\mathsf{N}'_\Xi) \subseteq D(g)$ and, for all $n \in D(\mathsf{N}'_\Xi)$, we have

$$g(n) \in D(\mathsf{N}_\Xi) \quad \text{and} \quad \mathsf{N}_\Xi(g(n)) = \mathsf{N}'_\Xi(n).$$

The definitions of $\mathsf{N}_\Xi \simeq_\mathrm{T} \mathsf{N}'_\Xi$, $\mathsf{N}_\Xi \prec_\mathrm{T} \mathsf{N}'_\Xi$, and $\mathsf{N}_\Xi \not\simeq_\mathrm{T} \mathsf{N}'_\Xi$ are analogous to the definitions of $\mathsf{N}_\Xi \simeq_\mu \mathsf{N}'_\Xi$, $\mathsf{N}_\Xi \prec_\mu \mathsf{N}'_\Xi$, and $\mathsf{N}_\Xi \not\simeq_\mu \mathsf{N}'_\Xi$. Further, note that we have

$$\mathsf{N}_\Xi \leq_\mathrm{T} \mathsf{N}'_\Xi \quad \Rightarrow \quad \mathsf{N}_\Xi \leq_\mu \mathsf{N}'_\Xi$$

for all numberings $\mathsf{N}_\Xi, \mathsf{N}'_\Xi : \mathbb{N} \supseteq \to \mathbb{N}$ (and all sets $\Xi_\mu$). This implication will become relevant in Theorems 1 and 3.

Definition 1 below will introduce the sets $\mathbb{R}_\mu$, $\Pi_2$, $\Sigma_2$, and $\Delta_2$, each of which is a special subset of $\mathbb{R}$. In the following, consider '$\Xi$' a placeholds symbol: We may consistently substitute either '$\mathbb{R}$', '$\Pi$', '$\Sigma$', or '$\Delta$' for '$\Xi$'.

Observe that Definition 1 will characterize the relevant computable objects through tuples of $\mu$-recursive functions. Albeit not evident, these characterizations are in line with the notion of modest sets. Fixing an *admissible enumeration* $(\check{\varphi}_n)_{n \in \mathbb{N}}$ of $\mu$-recursive functions [8, Chapter I, Section 3, p. 14ff], a $\mu$-recursive *pairing function* $\mathbf{p} : \mathbb{N} \times \mathbb{N} \to \mathbb{N}$ [11, Section 1.4, p. 12], and the projections $\mathbf{f}, \mathbf{s} : \mathbb{N} \to \mathbb{N}$ corresponding to the pairing function's inverse, it is possible to define $\mathsf{N}_\Xi : \mathbb{N} \supseteq \to \mathbb{N}$ s.t. for all $n \in D(\mathsf{N}_\Xi)$, we can retrieve the tuple of $\mu$-recursive functions that characterizes $\mathsf{N}_\Xi(n) \in \Xi_\mu$ (in the sense of Definition 1) from the realizer $n$. This way, the objects in $\Xi_\mu$ become accessible to computations themselves. Particularly, a sequence $(\xi_n)_{n \in \mathbb{N}} \subset \Xi_\mu$ is computable (in the sense of Definitions 1) if and only if there exists a $\mu$-recursive function $g : \mathbb{N} \to \mathbb{N}$ s.t. for all $n \in \mathbb{N}$, we have $\xi_n = \mathsf{N}_\Xi(g(n))$. For $N \in \mathbb{N}$, an $N$-ary $\mu$-recursive function $g : \mathbb{N}^N \to \mathbb{N}$ analogously defines an $N$-fold computable sequence.

The above principles analogously apply to Definition 2 in Section IV and Definition 4 in Section V: Both definitions ultimately characterize the relevant computable objects through $\mu$-recursive functions, and it is possible to define the associated numberings in a way that allows for retrieving these functions from each appropriate realizer. This way, Definitions 2 and 4 become consistent with the previously established notion of modest sets. Computable sequences in the sense of Definitions 2 and 4 exhibit equivalent definitions in terms of the relevant numberings, analogous to the above case. Then, for $N \in \mathbb{N}$, $N$-ary $\mu$-recursive functions define $N$-fold computable sequences as before.

Finally, note that as it is not of particular necessity to our purpose, we will not provide explicit details on the structure of the numberings we introduce in Definitions 1, 2, and 4.

**Definition 1** ($\mathbb{R}_\mu$, $\Pi_2$, $\Sigma_2$, and $\Delta_2$)**.** *In the following, we may consistently substitute '$\Pi_2$' and '$\limsup$', '$\Sigma_2$' and '$\liminf$', or '$\Delta_2$' and '$\lim$' for '$\Xi_\mu$' and '$\Theta$', respectively.*

- *We refer to a sequence $(x_n)_{n \in \mathbb{N}} \subset \mathbb{R}$ as* computable in $\mathbb{R}_\mu$ *if there exist (total) $\mu$-recursive functions $g : \mathbb{N} \times \mathbb{N} \to \{0,1\}$, $h_1, h_2, \nu : \mathbb{N} \times \mathbb{N} \to \mathbb{N}$, s.t. for all $n, m, M \in \mathbb{N}$, $m \geq \nu(n, M)$, we have*

$$\left| x_n - \frac{(-1)^{g(n,m)} \cdot h_1(n,m)}{1 + h_2(n,m)} \right| < \frac{1}{2^M}.$$

- *We refer to a sequence $(x_n)_{n\in\mathbb{N}} \subset \mathbb{R}$ as computable in $\Xi_\mu$ if there exist (total) $\mu$-recursive functions $g : \mathbb{N} \times \mathbb{N} \to \{0,1\}$, $h_1, h_2 : \mathbb{N} \times \mathbb{N} \to \mathbb{N}$, s.t. for all $n \in \mathbb{N}$, we have*

$$x_n = \Theta_{m\to\infty} \frac{(-1)^{g(n,m)} \cdot h_1(n,m)}{1 + h_2(n,m)}.$$

*The set of* computable numbers, *denoted by $\mathbb{R}_\mu$, is the smallest subset of $\mathbb{R}$ s.t. $\{x_n\}_{n\in\mathbb{N}} \subset \mathbb{R}_\mu$ for all sequences $(x_n)_{n\in\mathbb{N}}$ that are computable in $\mathbb{R}_\mu$. The definitions of $\Pi_2$, $\Sigma_2$, and $\Delta_2$ are analogous. We denote the corresponding numberings by $\mathsf{N}_\mathbb{R}$, $\mathsf{N}_\Pi$, $\mathsf{N}_\Sigma$, and $\mathsf{N}_\Delta$, respectively.*

The set $\mathbb{R}_\mu$ consists precisely of those real numbers we can algorithmically approximate up to any desired accuracy, such as $e$, $\pi$, or $\sqrt{n}$ for any $n \in \mathbb{N}$. Further $\mathbb{R}_\mu$, is closed with respect to addition and multiplication.

We have $\mathbb{R}_\mu \subset \Delta_2 = (\Pi_2 \cap \Sigma_2) \subset (\Pi_2 \cup \Sigma_2) \subset \mathbb{R}$. Within the *arithmetical hierarchy of real numbers* [17], the set $\mathbb{R}_\mu$ coincides with the set $\Delta_1 = (\Pi_1 \cup \Sigma_1)$. The arithmetical hierarchy of real numbers is a classification of (definable) real numbers or, depending on the context, real-valued functions in terms of their structural complexity; Theorems 2 and 4 will place the *fractional degree of smoothness* (c.f. Definition 3) and the *fractional degree of decay* (c.f. Definition 5) into this classification system. Previously, we have established such characterizations for the *actual bandwidth* of bandlimited functions [18], the *radius of convergence* of the $\mathcal{Z}$-transform of time-computable signals [19], and the *separation distance* of support sets of continuous functions [20].

From a practical perspective, the arithmetical hierarchy of real numbers is primarily relevant for its role in complexity theory. It is strongly related to the *Kleene–Mostowski hierarchy*. Upon considering specific subsets of $\mu$-recursive functions, the Kleene–Mostowski hierarchy becomes the *PSPACE* complexity hierarchy. In this sense, the complexity classes *NP* and *co-NP* are analogous to, respectively, the sets $\Sigma_1$ and $\Pi_1$ within the arithmetical hierarchy of real numbers.

In the subsequent Sections IV and V, we will apply the theory of computable analysis to the (real) Wiener Algebra $\mathbb{A}^0$ of periodic functions $f : \mathcal{I} \to \mathbb{R}$ and the $\ell^p$-spaces of asymptotically vanishing sequences, respectively. Do to their relevance in the modeling of physical processes on the one hand and, as we will demonstrate, their specific mathematical structure, they proved an excellent example of an application of the algorithmic-information concept. We will require the following technical lemma in establishing our theory.

**Lemma 1.** *Let $(x_{n,m})_{n,m\in\mathbb{N}}$ and $y$ be computable in $\mathbb{R}_\mu$. There exists a $\mu$-recursive function $h : \mathbb{N} \supseteq\to \mathbb{N}$ with domain $D(h) = \{n \in \mathbb{N} \mid \exists m \in \mathbb{N}.\ x_{n,m} < y\}$ s.t. for all $n \in D(h)$, we have $x_{n,h(n)} < y$.* □

In the terminology of recursion theory, Lemma 1 states the *semi-decidability* of the set $\mathbb{R}_{<y} \cap \mathbb{R}_\mu := \{x \in \mathbb{R}_\mu : x < y\}$, $y \in \mathbb{R}_\mu$. For details, we again refer to [8].

## IV. THE ALGORITHMIC-INFORMATION HIERARCHY OF PERIODIC FUNCTIONS

As indicated in Sections I and III, we subsequently establish an *effective* (i.e., computable) *fractional calculus* for the *Wiener algebra* $\mathbb{A}^0$ of *periodic functions* $f : \mathcal{I} \to \mathbb{R}$. Fractional calculus and periodic functions are vital to the mathematical modeling of physical processes. For example, fractional calculus is pivotal to *fractional PID control*, but also occurs in a range of other remarkable settings throughout general control theory [21]–[23]. Regarding our work, *repetitive control* [24], [25] is arguably the most relevant application of periodic functions, as it makes explicit use of the *internal-model principle* (c.f. Section I).

In the following, we denote the set of non-negative rational numbers by $\mathbb{Q}_{\geqslant 0}$ and, for $\omega \in \mathbb{Q}_{\geqslant 0}$, define

$$(\mathfrak{e}_{\omega,n})_{n\in\mathbb{N}} \ : \ \mathfrak{e}_{\omega,n} = \begin{cases} 1, & \text{if } \omega = 0, \\ \exp_n(\omega), & \text{if } \omega > 0. \end{cases}$$

The sequence $(\mathfrak{e}_{\omega,n})_{n\in\mathbb{N}}$ is $\mathbb{R}_\mu$-computable. Further, note that $\mathfrak{e}_{0,0} = 1$, whereas "$\exp_0(0)$" is *undefined*. Finally, for $n \in \mathbb{N}$, we denote the set of functions $f : \mathcal{I} \to \mathbb{R}$ whose $n$th *derivative* is *continuous* by $\mathcal{C}^n(\mathcal{I}, \mathbb{R})$.

**Definition 2** ($\mathbb{A}_\mu^\omega$). *We call a sequence $(f_n)_{n\in\mathbb{N}} : \mathcal{I} \to \mathbb{R}$ computable in $\mathbb{A}_\mu^\omega$, $\omega \in \mathbb{Q}_{\geqslant 0}$, if there exist $(x_{n,m})_{n,m\in\mathbb{N}}$ and $\nu : \mathbb{N} \times \mathbb{N} \to \mathbb{N}\backslash\{0\}$ s.t.*

- *for all $n \in \mathbb{N}$, $t \in \mathcal{I}$, the function $f_n$ satisfies $f_n(t) = \ldots$
$\ldots x_{n,0} + \sum_{m=1}^{\infty} x_{n,2m}\cos_m(t) + x_{n,2m-1}\sin_m(t)$;*
- *for all $n, M \in \mathbb{N}$ and with $\jmath$ denoting the imaginary unit, we have $1/2^M > \sum_{m=\nu(n,M)}^{\infty} \mathfrak{e}_{\omega,m}|x_{n,2m} + \jmath x_{n,2m-1}|$;*
- *$(x_{n,m})_{n,m\in\mathbb{N}}$ is a computable sequence in $\mathbb{R}_\mu$ and $\nu$ is a (total) $\mu$-recursive function.*

*The set of $\mathbb{A}_\mu^\omega$-computable functions, denoted by $\mathbb{A}_\mu^\omega$, is the smallest subset of $\mathcal{C}^0(\mathcal{I}, \mathbb{R})$ s.t. $\{f_n\}_{n\in\mathbb{N}} \subset \mathbb{A}_\mu^\omega$ for all sequences $(f_n)_{n\in\mathbb{N}}$ that are computable in $\mathbb{A}_\mu^\omega$. We denote the relevant numbering by $\mathsf{N}_{\mathbb{A},\omega}$.*

We call $\mathbb{A}_\mu^0$ the *computable (real) Wiener algebra* of (computable) *periodic functions* $f : \mathcal{I} \to \mathbb{R}$.

Consider $\omega \in \mathbb{Q}_{\geqslant 0}$ arbitrary and let $(x_n)_{n\in\mathbb{N}} \subset \mathbb{R}_\mu$ characterize $f \in \mathbb{A}_\mu^\omega$ in the sense of Definition 2. Define $\|\cdot\|_\omega^\sim : \mathbb{A}_\mu^\omega \to \mathbb{R}_\mu$ through

$$\|f\|_\omega^\sim := \max_{\tau\in\{0,\omega\}} \left(\mathfrak{e}_{\tau,0}|x_0| + \sum_{n=1}^{\infty}\mathfrak{e}_{\tau,n}|x_{2n} + \jmath x_{2n-1}|\right).$$

Then, $\|\cdot\|_\omega^\sim$ defines a *norm* in $\mathbb{A}_\mu^\omega$, and the pair $(\mathbb{A}_\mu^\omega, \|\cdot\|_\omega^\sim)$ satisfies the following *structural properties*.

- Let $(f_{1,n})_{n\in\mathbb{N}}, (f_{2,n})_{n\in\mathbb{N}} \subset \mathbb{A}_\mu^\omega$, and $(x_n)_{n\in\mathbb{N}} \subset \mathbb{R}_\mu$ be any computable sequences (in $\mathbb{A}_\mu^\omega$ and $\mathbb{R}_\mu$, respectively). Then, $(f_n)_{n\in\mathbb{N}} : f_n = x_n(f_{1,n} + f_{2,n})$ is computable in $\mathbb{A}_\mu^\omega$.
- Let $(f_{n,m})_{n,m\in\mathbb{N}}$ be computable in $\mathbb{A}_\mu^\omega$ and assume there exists a (total) $\mu$-recursive function $\nu : \mathbb{N} \times \mathbb{N} \to \mathbb{N}$ s.t.

$$\|f_{n,\nu(n,m)+l} - f_{n,\nu(n,m)+k}\|_\omega^\sim < 2^{-m}$$

for all $n, m, l, k \in \mathbb{N}$. Then, $(f_{n,*})_{n\in\mathbb{N}} : f_{n,*} = \lim_{m\to\infty} f_{n,m}$ is computable in $\mathbb{A}_\mu^\omega$ as well. We say that $(f_{n,m})_{n,m\in\mathbb{N}}$ converges *effectively* to $(f_{n,*})_{n\in\mathbb{N}}$ for $m \to \infty$.

- Let $(f_n)_{n\in\mathbb{N}}$ be computable in $\mathbb{A}_\mu^\omega$. Then, the sequence $(x_n)_{n\in\mathbb{N}} : x_n = \|f_n\|_{\widetilde{\omega}}$ is computable in $\mathbb{R}_\mu$.

In other words, $\mathbb{A}_\mu^\omega$ is effectively closed w. r. to addition, (scalar) multiplication, and (effective) $\|\cdot\|_{\widetilde{\omega}}$-convergence. Further, if $(f_n)_{n\in\mathbb{N}}$ is computable in $\mathbb{A}_\mu^\omega$, we can always compute the corresponding values of $\|\cdot\|_{\widetilde{\omega}}$. Accordingly, the pair $(\mathbb{A}_\mu^\omega, \|\cdot\|_{\widetilde{\omega}})$ forms an *effective Banach Space*.

Given any effective Banach Space $\mathbb{B}_\mu$ and $N, M \in \mathbb{N}$, consider the binary relation

$$\hat{v} \rightleftarrows_\mathbb{B} v := \big( (\|\hat{v} - v\|_\mathbb{B} \leqslant 2^{-N}) \Rightarrow (\|v\|_\mathbb{B} < 2^{-M}) \big), \quad (4)$$

which is a modified and simpler variant of (1). Specifically, (4) aims to ensure $v$ *itself* (rather than the differences of its subvectors) to not *exceed* (rather than fall below) a critical value $1/2^M$ in magnitude. With $q = 1/2^M - 1/2^N$ and by means of the triangle inequality, we then obtain

$$D(\mathsf{GO}_\rightleftarrows) = \{\hat{v} \in \mathbb{B}_\mu \mid \|\hat{v}\|_\mathbb{B} < q\}.$$

Due to Lemma 1, we can achieve GO integrity to the least trivial extent for properties of the form (4) in any effective Banach Space $\mathbb{B}_\mu$ – provided we employ a suitable numbering. That is, provided the realizers in the chosen numbering contain the relevant algorithmic information. Theorem 1 below will formalize this observation.

Consider $n \in \mathbb{N}$ and $\tau, \omega \in \mathbb{Q}_{\geqslant 0}$ s.t. $\tau < \omega < 1$, otherwise arbitrary. Then, we have

$$\mathcal{C}^{n+1}(\mathcal{I}, \mathbb{R}) \subset \mathbb{A}_\mu^{n+\omega} \subset \mathbb{A}_\mu^{n+\tau} \subseteq \mathcal{C}^n(\mathcal{I}, \mathbb{R}).$$

In this regard, the sets $\mathbb{A}_\mu^\omega$, $\omega \in \mathbb{Q}_{\geqslant 0}$, generalize the concept of higher-order continuous differentiability to the non-integer domain. Particularly, let

$$\mathfrak{D}^\tau : \mathbb{A}_\mu^\omega \to \mathbb{A}_\mu^{\omega-\tau}, \quad f \mapsto \mathfrak{D}^\tau f, \quad (5)$$

be a *fractional differential operator* of *degree* $\tau \in \mathbb{Q}_{\geqslant 0}$ (c.f., for example [1]), where $\tau \leqslant \omega$. Then, $\mathfrak{D}^\tau f$ is a well-defined element of $\mathbb{A}_\mu^{\omega-\tau}$ if and only if we have $f \in \mathbb{A}_\mu^\omega$. In other words, $f$ must be "sufficiently smooth", which motivates the subsequent Definition 3.

**Definition 3** ($\Omega$). *Consider $f \in \mathbb{A}_\mu^0$ and let $(x_n)_{n\in\mathbb{N}} \subset \mathbb{R}_\mu$ characterize $f$ in the sense of Definition 2. With $(y_n)_{n\in\mathbb{N}\setminus\{0\}} : y_n = |x_{2n} + \jmath x_{2n-1}|$, define*

$$\Omega(f) := \sup\Big\{\omega \in \mathbb{Q}_{\geqslant 0} \,\Big|\, \exists q \in \mathbb{Q}_{\geqslant 0}. \forall N \in \mathbb{N}.\ q > \sum_{n=1}^N \mathfrak{e}_{\omega,n} y_n \Big\}.$$

*We call $\Omega(f)$ the* fractional degree of smoothness *of $f$.*

Note that we may have $\Omega(f) = \infty$ if the corresponding sequence $(x_n)_{n\in\mathbb{N}}$ decays sufficiently fast for $n \to \infty$.

Consider $f \in \mathbb{A}_\mu^0$ and $\omega \in \mathbb{Q}_{\geqslant 0}$, $\omega < \Omega(f)$. Then, $\|f\|_{\widetilde{\omega}}$ is well-defined. However, it is not evident whether we necessarily have $f \in \mathbb{A}_\mu^\omega$. Theorem 1 below will prove that this is indeed the case: If $\omega < \Omega(f)$, there always exists $n \in \mathbb{N}$ s.t. $f = \mathsf{N}_{\mathbb{A},\omega}(n)$. Thus, each of the spaces $\mathbb{A}_\mu^\omega$, $\omega \in \{\tau \in \mathbb{Q}_{\geqslant 0} | \tau < \Omega(f)\}$, induces a characteristic set of realizers of $f$, motivating the definition

$$\mathsf{N}_{\mathbb{A},\tau|\omega} : \mathbb{N} \supseteq \to \mathbb{A}_\mu^\omega, \quad n \mapsto \mathsf{N}_{\mathbb{A},\tau|\omega}(n) := \mathsf{N}_{\mathbb{A},\tau}(n), \text{ with}$$
$$D(\mathsf{N}_{\mathbb{A},\tau|\omega}) := \{n \in D(\mathsf{N}_{\mathbb{A},\tau}) : \mathsf{N}_{\mathbb{A},\tau}(n) \in \mathbb{A}_\mu^\omega\}$$

for all $\tau, \omega \in \mathbb{Q}_{\geqslant 0}$, $\tau < \omega$. In essence, $\mathsf{N}_{\mathbb{A},\tau|\omega}$ is the *restriction* of $\mathsf{N}_{\mathbb{A},\tau}$ to realizers of functions in $\mathbb{A}_\mu^\omega$. Due to the inclusion $\mathbb{A}_\mu^\omega \subset \mathbb{A}_\mu^\tau$, both $\mathsf{N}_{\mathbb{A},\tau|\omega}$ and $\mathsf{N}_{\mathbb{A},\omega}$ are numberings for the (same) set $\mathbb{A}_\mu^\omega$.

**Theorem 1.** *Consider $\tau, \omega, q \in \mathbb{Q}_{\geqslant 0}$ s.t. $\tau < \omega$ and $0 < q < 1$.*
1) *Let $f \in \mathbb{A}_\mu^\tau$ be arbitrary and assume we have $\omega < \Omega(f)$. Then, we also have $f \in \mathbb{A}_\mu^\omega$.*
2) *Let $f \in \mathbb{A}_\mu^\omega$ and $m \in \mathbb{N}$ be arbitrary and assume we have $\mathsf{N}_{\mathbb{A},\omega}(m) = f$. Then, we also have $\mathsf{N}_{\mathbb{A},\tau}(m) = f$.*
3) *There exists an $\mathbb{A}_\mu^\tau$-computable sequence $(f_{n,*})_{n\in\mathbb{N}} \subset \mathbb{A}_\mu^\omega$ s.t. no $\mu$-recursive function $g : \mathbb{N} \supseteq \to \mathbb{N}$ satisfies*

$$D(g) = \{n \in \mathbb{N} \mid \|f_{n,*}\|_{\widetilde{\omega}} < q\}.$$

*Thus, $(x_n)_{n\in\mathbb{N}} : x_n = \|f_{n,*}\|_{\widetilde{\omega}}$ is not an $\mathbb{R}_\mu$-computable and $(f_{n,*})_{n\in\mathbb{N}}$ not an $\mathbb{A}_\mu^\omega$-computable sequence.* [1]

*Consequently (as follows from Statements 2 and 3), we have $\mathsf{N}_{\mathbb{A},\tau|\omega} \prec_\mathrm{T} \mathsf{N}_{\mathbb{A},\omega}$ and $\mathsf{N}_{\mathbb{A},\tau|\omega} \prec_\mu \mathsf{N}_{\mathbb{A},\omega}$.*

*Proof Sketch (Theorem 1).*

*Statement 1.* Let $(x_n)_{n\in\mathbb{N}} \subset \mathbb{R}_\mu$ characterize $f$ in the sense of Definition 2. We must demonstrate the existence of a corresponding $\mu$-recursive modulus of convergence $\nu : \mathbb{N} \to \mathbb{N}\setminus\{0\}$. To this end, consider

$$\sigma, q \in \mathbb{Q}_{\geqslant 0} \quad \text{s.t.} \quad \omega < \sigma < \Omega(f) \text{ and } q \geqslant \|f\|_{\widetilde{\sigma}}.$$

Accordingly, with $p = \sigma - \omega$ and $(y_n)_{n\in\mathbb{N}\setminus\{0\}} : y_n = |x_{2n} + \jmath x_{2n-1}|$, we have

$$\sum_{n=M}^\infty \mathfrak{e}_{\omega,n} y_n \leqslant \mathfrak{e}_{p,M}^{-1} \sum_{n=M}^\infty \mathfrak{e}_{\sigma,n} y_n \leqslant q \mathfrak{e}_{p,M}^{-1}$$

for all $M \in \mathbb{N}\setminus\{0\}$. Furthermore, the sequence $(z_{n,m})_{n,m\in\mathbb{N}} : z_{n,m} = q\mathfrak{e}_{p,m+1}^{-1} - 1/2^n$ is computable in $\mathbb{R}_\mu$. By Lemma 1, there exists a $\mu$-recursive function $h : \mathbb{N} \to \mathbb{N}$ s.t. for all $n \in \mathbb{N}$, we have $z_{n,h(n)} < 0$. Defining $\nu : \mathbb{N} \to \mathbb{N}\setminus\{0\}$ through $\nu(n) := h(n) + 1$ then concludes the proof.

*Statement 2.* For $m \in D(\mathsf{N}_{\mathbb{A},\omega})$, let $((x_n)_{n\in\mathbb{N}}, \nu)$ be the unique corresponding pair that characterizes $\mathsf{N}_{\mathbb{A},\omega}(m) \in \mathbb{A}_\mu^\omega$ in the sense of Definition 2. For $(y_n)_{n\in\mathbb{N}\setminus\{0\}} : y_n = |x_{2n} + \jmath x_{2n-1}|$ and all $M \in \mathbb{N}$, the pair $((x_n)_{n\in\mathbb{N}}, \nu)$ satisfies

$$\sum_{n=\nu(M)}^\infty \mathfrak{e}_{\tau,n} y_n \leqslant \sum_{n=\nu(M)}^\infty \mathfrak{e}_{\omega,n} y_n \leqslant 2^{-M}.$$

Hence, we must have $m \in D(\mathsf{N}_{\mathbb{A},\tau})$. Since we required $m \in D(\mathsf{N}_{\mathbb{A},\omega})$, we must thus also have $\mathsf{N}_{\mathbb{A},\tau}(m) = \mathsf{N}_{\mathbb{A},\omega}(m)$.

*Statement 3.* Let $h : \mathbb{N} \to \mathbb{N}$ be any total $\mu$-recursive function s.t. $\mathcal{A} := \{h(n) | n \in \mathbb{N}\}$ is *not* recursive. We define

$$(\mathcal{A}_m)_{m\in\mathbb{N}} : \quad \mathcal{A}_m = \bigcup_{n=1}^m \{h(n-1)\},$$

$$(f_{n,m})_{n,m\in\mathbb{N}} : \quad f_{n,m} = \begin{cases} f_{n,m-1}, & \text{if } n \in \mathcal{A}_m, \\ \mathfrak{e}_{\omega,m}^{-1}\cos_m, & \text{otherwise.} \end{cases}$$

Note that $\mathcal{A}_0 = \cup_{n=1}^0 \{h(n-1)\} = \emptyset$ and $\mathcal{A} = \cup_{m=0}^\infty \mathcal{A}_m$. Furthermore, observe the following.

- The sequence $(f_{n,m})_{n,m\in\mathbb{N}}$ is computable in $\mathbb{A}_\mu^\omega$ and we have $\|f_{n,m}\|_{\widetilde{\omega}} = 1$ for all $n, m \in \mathbb{N}$.
- The sequence $(f_{n,m})_{n,m\in\mathbb{N}}$ is computable in $\mathbb{A}_\mu^\tau$ and we have $\|f_{n,m} - f_{n,m+l}\|_{\widetilde{\tau}} < \mathfrak{e}_{\omega-\tau,m}^{-1}$ for all $n, m \in \mathbb{N}$.

---

[1] For all $n \in \mathbb{N}$, we have $f_{n,*} \in \mathbb{A}_\mu^\omega$. However, $(f_{n,*})_{n\in\mathbb{N}}$ is not *computable* (as a sequence) in $\mathbb{A}_\mu^\omega$.

As follows from the above, $(f_{n,*})_{n\in\mathbb{N}} : f_{n,*} = \lim_{m\to\infty} f_{n,m}$ (where the limit is with respect to $\|\cdot\|_{\widetilde{\tau}}$) is a computable sequence in $\mathbb{A}_\mu^\tau$ and, with $\mathcal{A}^c = \mathbb{N}\setminus\mathcal{A}$, satisfies the following.

- For all $n\in\mathcal{A}$, there exists $m\in\mathbb{N}$ s.t. $f_{n,*} = f_{n,m} \in \mathbb{A}_\mu^\omega$, and we have $\|f_{n,*}\|_{\widetilde{\omega}} = 1$ accordingly.
- For all $n\in\mathcal{A}^c$, we have $f_{n,*} = \mathbf{0} \in \mathbb{A}_\mu^\omega$ and thus $\|f_{n,*}\|_{\widetilde{\omega}} = 0$.

Consequently, we have $\mathcal{A}^c = \{n\in\mathbb{N}\,|\,\|f_{n,*}\|_{\widetilde{\omega}} - 1/2 < 0\}$. As we chose $\mathcal{A}$ to not be recursive, $\mathcal{A}^c$ cannot be the domain of any $\mu$-recursive function and, as then follows from Lemma 1 by contradiction, $(x_n)_{n\in\mathbb{N}} : x_n = \|f_{n,*}\|_{\widetilde{\omega}}$ cannot be an $\mathbb{R}_\mu$-computable sequence. The remainder of Statement 3 is then, again by contradiction, due to $\mathbb{A}_\mu^\omega$ being an effective Banach space. □

Note that we must interpret Statement 1 of Theorem 1 in terms of mere *mathematical existence*: Given $f\in\mathbb{A}_\mu^\tau$, there *exists* a number $n\in\mathbb{N}$ s.t. $f = \mathsf{N}_{\mathbb{A},\omega}(n)$. However, as demonstrated by Statement 3 of Theorem 1, algorithmically constructing $n\in\mathbb{N}$ s.t. $f = \mathsf{N}_{\mathbb{A},\omega}(n)$ from $m\in\mathbb{N}$ satisfying $f = \mathsf{N}_{\mathbb{A},\tau}(m)$ is impossible. Observe that for $\omega,\tau\in\mathbb{Q}_{\geq 0}$, $\tau\leq\omega$, we have

$$D(\mathsf{N}_{\mathbb{A},0}) \supseteq D(\mathsf{N}_{\mathbb{A},\tau}) \supseteq D(\mathsf{N}_{\mathbb{A},\tau|\omega}) \supseteq D(\mathsf{N}_{\mathbb{A},\omega}) \quad (6)$$

and, by Statement 2 of Theorem 1, $\mathsf{N}_{\mathbb{A},\omega}(m) = \mathsf{N}_{\mathbb{A},\tau}(m)$, for all $m\in D(\mathsf{N}_{\mathbb{A},\omega})$. If $f = \mathsf{N}_{\mathbb{A},0}(n) = \mathsf{N}_{\mathbb{A},0}(m)$ for some $f\in\mathbb{A}_\mu^0$, $n,m\in\mathbb{N}$, both $n$ and $m$ determine $f$ unambiguously in a mathematical sense. On an abstract logical level, both $n$ and $m$ thus contain *all* possible information about $f$. However, if we have $n\in D(\mathsf{N}_{\mathbb{A},\omega})$ but $m\in\mathbb{N}\setminus D(\mathsf{N}_{\mathbb{A},\omega})$ for some $\omega\in\mathbb{Q}_{\geq 0}$, $\omega > 0$, the realizer $n$ contains more *algorithmic* information (about $f$) than the realizer $m$.

As a consequence of Statement 3 in Theorem 1, computing the $\|\cdot\|_{\widetilde{\omega}}$-norm of functions $f\in\mathbb{A}_\mu^0$ for $\omega\in\mathbb{Q}_{\geq 0}$, $\omega < \Omega(f)$ is possible only if we choose a numbering that contains a sufficient amount of algorithmic information. Statement 3 of Theorem 1 thus provides a direct link between the GO integrity for properties of the form 4 and the algorithmic information of numberings for the subsets $\mathbb{A}_\mu^\omega$, $\omega\in\mathbb{Q}_{\geq 0}$, of the computable real Wiener algebra $\mathbb{A}_\mu^0$. A similar relationship holds true for fractional differential operators of the form (5). To compute a fractional differential operator of order $\omega\in\mathbb{Q}_{\geq 0}$, we must represent its argument through $\mathsf{N}_{\mathbb{A},\omega}$ or higher.

Finally, we classify the fractional degree of smoothness – interpreted as a function $\Omega:\mathbb{A}_\mu^0\to\mathbb{R}$, $f\mapsto\Omega(f)$ – within the arithmetical hierarchy of real numbers.

**Theorem 2.**

1) Let $(f_n)_{n\in\mathbb{N}}$ be any $\mathbb{A}_\mu^0$-computable sequence. If the sequence $(x_n)_{n\in\mathbb{N}} : x_n = \Omega(f_n)$ is well-defined (that is, if $\Omega(f_n) < \infty$ for all $n\in\mathbb{N}$), it is $\Pi_2$-computable.
2) Let $(x_n)_{n\in\mathbb{N}}$ be any $\Pi_2$-computable sequence and consider $\tau\in\mathbb{Q}$, $\tau > 0$. There exists an $\mathbb{A}_\mu^0$-computable sequence $(f_n)_{n\in\mathbb{N}}$ s.t. for all $n\in\mathbb{N}$, we have $\Omega(f_n) = \max\{\tau, x_n\}$.

*Proof Sketch (Theorem 2).*

*Statement 1.* Let $(y_{n,m})_{n,m\in\mathbb{N}}$ determine $(f_n)_{n\in\mathbb{N}}$ in the sense of Definition 2 and define[2]

$(f_{n,m})_{n,m\in\mathbb{N}} : f_{n,m} = y_{n,0} + \sum_{l=1}^{m} y_{n,2l}\cos_l + y_{n,2l-1}\sin_l$,

$(\mathcal{A}_{n,m})_{n,m\in\mathbb{N}} : \mathcal{A}_{n,m} = \{\omega\in\mathbb{Q}_{\geq 0}\,|\,\exists M\in\mathbb{N}.\ m < \|f_{n,M}\|_{\widetilde{\omega}}\}$.

The sequence $(\mathcal{A}_{n,m})_{n,m\in\mathbb{N}}$ is a *computable sequence of recursively enumerable sets* in the following sense: There exist total $\mu$-recursive functions $g,h:\mathbb{N}^3\to\mathbb{N}$ s.t.

$$\mathcal{A}_{n,m} = \left\{\tfrac{g(n,m,l)}{1+h(n,m,l)}\,\Big|\,l\in\mathbb{N}\right\}.$$

for all $n,m\in\mathbb{N}$. As follows from the construction of $(f_{n,m})_{n,m\in\mathbb{N}}$ and $(\mathcal{A}_{n,m})_{n,m\in\mathbb{N}}$, we thus have

$$\Omega(f_n) = \sup_{m\in\mathbb{N}}\inf_{l\in\mathbb{N}}\frac{g(n,m,l)}{1+h(n,m,l)}$$

for all $n\in\mathbb{N}$. Then, [17, Lemma 3.2, p. 55] implies the claim.

*Statement 2.* Define the total $\mu$-recursive function $\lambda:\mathbb{N}\to\mathbb{N}$ through

$$\lambda(m) := \max\left\{l\in\mathbb{N}\,\big|\,l\leq\sum_{n=1}^{m}1/n\right\}$$

and observe that as a consequence of Definition 1, there exists an $\mathbb{R}_\mu$-computable sequence $(x_{n,m})_{n,m\in\mathbb{N}}\subset\mathbb{Q}$ s.t. for all $n\in\mathbb{N}$, we have $x_n = \liminf_{m\to\infty} x_{n,m}$. Then,[2]

$(y_{n,m})_{n,m\in\mathbb{N}} : y_{n,m} = \max\{\tau, x_{n,\lambda(m)}\} + 1$ and

$(f_{n,m})_{n,m\in\mathbb{N}} : f_{n,m} = \sum_{l=1}^{m+1}\sin_l\exp_l(-y_{n,l-1})$,

define an $\mathbb{R}_\mu$-computable and, with $\omega\in\mathbb{Q}_{\geq 0}$ arbitrary, an $\mathbb{A}_\mu^\omega$-computable sequence, respectively. Particularly, we have

$$\|f_{n,m} - f_{n,l}\|_{\widetilde{\omega}} = \sum_{k=l+2}^{m+1}\exp_k\big(-(y_{n,k-1}-\omega)\big)$$
$$\leq \sum_{k=l+2}^{m+1}\exp_k\big(\omega-(1+\tau)\big)$$

for all $n,m,l\in\mathbb{N}$, $l\geq m$. Moreover, choose $\omega,q\in\mathbb{Q}_{>0}$ s.t. $\omega < \tau$ and $q > \zeta(1+\tau-\omega)$ (where $\zeta:\mathbb{C}\supseteq\to\mathbb{C}$ denotes the *Riemann zeta function*), in which case we have

$$\|f_{n,m} - f_{n,l}\|_{\widetilde{0}} \leq \sum_{k=l+2}^{m+1}\exp_k\big(\omega-\omega-(1+\tau)\big)$$
$$\leq \mathfrak{e}_{\omega,l+2}^{-1}\zeta(1+\tau-\omega) \leq q\mathfrak{e}_{\omega,l+2}^{-1} \quad (7)$$

for all $n,m,l\in\mathbb{N}$, $l\geq m$. Consequently, with respect to $\|\cdot\|_{\widetilde{0}}$, the sequence

$(f_{n,*})_{n\in\mathbb{N}} : f_{n,*} = \lim_{m\to\infty} f_{n,m}$

is well-defined. Furthermore, due to $\mathbb{A}_\mu^0$ being an effective Banach space, (7) implies the computability of $(f_{n,*})_{n\in\mathbb{N}}$ in $\mathbb{A}_\mu^0$. Lastly, choose $n\in\mathbb{N}$ and $\omega\in\mathbb{Q}_{\geq 0}$ arbitrarily. If $\omega < \Omega(f_{n,*})$, we have

$$\|f_{n,*}\|_{\widetilde{\omega}} = \sum_{m=0}^{\infty}\exp_{m+1}(\omega - y_{n,m}). \quad (8)$$

On the other hand, if $\omega > \Omega(f_{n,*})$, the right-hand side of (8) diverges. By comparison to the *harmonic series*, we obtain $\Omega(f_{n,*}) = \max\{\tau, x_n\}$. Since we chose $n\in\mathbb{N}$ arbitrarily, Statement 2 follows. □

---

[2] Note that $(f_{n,m})_{n,m\in\mathbb{N}}$ consists entirely of finite linear combinations of sin- (and cos-) functions, so that we have $(f_{n,m})_{n,m\in\mathbb{N}}\subset\mathbb{A}_\mu^\omega$ for all $\omega\in\mathbb{Q}_{\geq 0}$ (in principle, we have used the same line of reasoning for the sequence $(f_{n,m})_{n,m\in\mathbb{N}}$ in the proof sketch of Theorem 1, Statement 3).

As follows from Theorem 2, $\Omega$ is $\Pi_2$-*complete*, i.e., it is structurally *exactly* as complex as all other $\Pi_2$-problems. Note that this immediately implies the *uncomputability* of $\Omega$ as a function, as $\Omega(f)$ will, for suitable $f \in \mathbb{A}_\mu^0$, attain *uncomputable* values.

## V. The Algorithmic-Information Hierarchy of Asymptotically Vanishing Sequences

The theory of functions in $\mathbb{A}_\mu^\omega$, $\omega \in \mathbb{Q}_{\geqslant 0}$, is, in essence, a theory of sequences $(x_n)_{n \in \mathbb{N}}$ that converge to zero sufficiently *fast* for $n \to \infty$. Particularly, a sequence's rate of convergence is determined by the corresponding function's fractional degree of smoothness. Outside the context of periodic functions, the theory of $\ell^p$-functions, $p \in \mathbb{Q}_{\geqslant 1}^\infty := \{q \in \mathbb{Q} | q \geqslant 1\} \cup \{\infty\}$, leads to a notable "dual"-problem of *slowly* converging sequences. Particularly, the set $\ell^p$ consists of those functions $f : \mathbb{N} \to \mathbb{C}$ that satisfy $\lim_{n \to \infty} f(n) = 0$ and whose $\ell^p$-norm $\|f\|_p$ is well-defined (i.e., finite), where

$$\|f\|_p := \begin{cases} \left(\sum_{n=0}^\infty |f(n)|^p\right)^{1/p}, & \text{if } p < \infty, \\ \sup_{n \in \mathbb{N}} |f(n)|, & \text{otherwise.} \end{cases}$$

With respect to (point-wise) addition and (point-wise) scalar multiplication, $\ell^p$ forms a Banach space.

Asymptotically vanishing sequences form a cornerstone of classical digital signal processing. In addition, $\ell^p$-spaces are structurally related to *modulation spaces*, which form the mathematical basis of *time-frequency analysis* [26], [27].

**Definition 4** ($\ell_\mu^p$). *We call a sequence* $(f_n)_{n \in \mathbb{N}} : \mathbb{N} \to \mathbb{C}$ *computable in* $\ell_\mu^p$, $p \in \mathbb{Q}_{\geqslant 1}^\infty$, *if there exist* $(x_{n,m})_{n,m \in \mathbb{N}}$ *and* $\nu : \mathbb{N} \times \mathbb{N} \to \mathbb{N} \setminus \{0\}$ *s.t. (with $\jmath$ denoting the imaginary unit)*
- *we have* $f_n(m) = x_{n,2m} + \jmath x_{n,2m+1}$ *for all* $n, m \in \mathbb{N}$;
- *if* $p < \infty$, *we have* $1/2^M > \sum_{m=\nu(n,M)}^\infty |x_{n,2m} + \jmath x_{n,2m+1}|^p$ *for all* $n, M \in \mathbb{N}$; *otherwise (if* $p = \infty$*), we have* $1/2^M \geqslant |x_{n,2m} + \jmath x_{n,2m+1}|$ *for all* $n, m, M \in \mathbb{N}$, $m \geqslant \nu(n,M)$;
- $(x_{n,m})_{n,m \in \mathbb{N}}$ *is a computable sequence in* $\mathbb{R}_\mu$ *and* $\nu$ *is a (total) $\mu$-recursive function.*

*The set of $\ell_\mu^p$-computable functions, denoted by $\ell_\mu^p$, is the smallest subset of $\ell^p$ s.t. $\{f_n\}_{n \in \mathbb{N}} \subset \ell_\mu^p$ for all sequences $(f_n)_{n \in \mathbb{N}}$ that are computable in $\ell_\mu^p$. We denote the relevant numbering by $\mathsf{N}_{\ell,p}$.*

For $p \in \mathbb{Q}_{\geqslant 1}^\infty$, the pair $(\ell_\mu^p, \|\cdot\|_p)$ forms an effective Banach space. In this regard, its structural properties (c.f. Section IV) are analogous to $(\mathbb{A}_\mu^\omega, \|\cdot\|_{\widetilde{\omega}})$, $\omega \in \mathbb{Q}_{\geqslant 0}$. Further, observe that $\mathbb{A}_\mu^0$ and $\ell_\mu^1$ are naturally isometric. Thus, we can identify $\mathbb{A}_\mu^\omega$ with a subset of $\ell_\mu^1$. With some abuse of notation, we obtain the chain of inclusions $\mathbb{A}_\mu^\omega \subseteq \mathbb{A}_\mu^0 \simeq \ell_\mu^1 \subseteq \ell_\mu^p \subset \ell_\mu^\infty$. Subsequently, we will introduce the *fractional degree of decay*, which is the $\ell^p$-analogon of the fractional degree of smoothness.

**Definition 5** ($\Psi$). *Consider* $f \in \ell_\mu^\infty$ *and let* $(x_n)_{n \in \mathbb{N}} \subset \mathbb{R}_\mu$ *characterize $f$ in the sense of Definition 4. With* $(y_n)_{n \in \mathbb{N}} : y_n = |x_{2n} + \jmath x_{2n+1}|$, *define*

$$\Psi(f) := \inf\left\{\omega \in \mathbb{Q}_{\geqslant 1} \,\middle|\, \exists q \in \mathbb{Q}_{\geqslant 0}. \forall N \in \mathbb{N}. \, q > \sum_{n=1}^N (y_n)^\omega\right\}.$$

*We call $\Psi(f)$ the fractional degree of decay of $f$.*

Note that we may have $\Psi(f) = \infty$ if the corresponding sequence $(x_n)_{n \in \mathbb{N}}$ decays sufficiently slow for $n \to \infty$. The definition of $\mathsf{N}_{\ell, \tau | \omega}$ for $\tau, \omega \in \mathbb{Q}_{\geqslant 1}$, $\tau > \omega$ is analogous to the definition of $\mathsf{N}_{\mathbb{A}, \tau | \omega}$ for $\tau, \omega \in \mathbb{Q}_{\geqslant 1}$, $\tau < \omega$.

**Theorem 3.** *Consider $\tau, \omega \in \mathbb{Q}_{\geqslant 1}$ and $q \in \mathbb{Q}_{\geqslant 0}$ s.t. $\tau > \omega$ and $1 > q > 0$.*
1) *Let $f \in \ell_\mu^\tau$ be arbitrary and assume we have $\omega > \Psi(f)$. Then, we also have $f \in \ell_\mu^\omega$.*
2) *Let $f \in \ell_\mu^\omega$ and $m \in \mathbb{N}$ be arbitrary and assume we have $\mathsf{N}_{\ell,\omega}(m) = f$. Then, we also have $\mathsf{N}_{\ell,\tau}(m) = f$.*
3) *There exists an $\ell_\mu^\tau$-computable sequence $(f_{n,*})_{n \in \mathbb{N}} \subset \ell_\mu^\omega$ s.t. no $\mu$-recursive function $g : \mathbb{N} \supseteq \to \mathbb{N}$ satisfies*

$$D(g) = \{n \in \mathbb{N} \mid \|f_{n,*}\|_\omega < q\}.$$

*Thus, $(x_n)_{n \in \mathbb{N}} : x_n = \|f_{n,*}\|_\omega$ is not an $\mathbb{R}_\mu$-computable and $(f_{n,*})_{n \in \mathbb{N}}$ not an $\ell_\mu^\omega$-computable sequence.* [3]

*Consequently (as follows from Statements 2 and 3), we have $\mathsf{N}_{\ell,\tau|\omega} <_T \mathsf{N}_{\ell,\omega}$ and $\mathsf{N}_{\ell,\tau|\omega} <_\mu \mathsf{N}_{\ell,\omega}$.* □

The interpretation of Theorem 3 follows along the same lines as the interpretation of Theorem 1. Note that we have

$$D(\mathsf{N}_{\ell,\omega}) \subseteq D(\mathsf{N}_{\ell,\tau|\omega}) \subseteq D(\mathsf{N}_{\ell,\tau}) \subset D(\mathsf{N}_{\ell,\infty})$$

for all $\omega, \tau \in \mathbb{Q}_{\geqslant 1}$, $\tau \geqslant \omega$. Note the reversal of direction as compared to (6). Specifically, for $\tau \geqslant \omega$, we have $D(\mathsf{N}_{\ell,\omega}) \subseteq D(\mathsf{N}_{\ell,\tau})$ but $D(\mathsf{N}_{\mathbb{A},\tau}) \subseteq D(\mathsf{N}_{\mathbb{A},\omega})$.

Finally, just as we did for the fractional degree of smoothness, we classify the fractional degree of decay – interpreted as a function $\Psi : \ell_\mu^\infty \to \mathbb{R} \cup \{\infty\}$, $f \mapsto \Psi(f)$ – within the arithmetical hierarchy of real numbers.

**Theorem 4.**
1) *Let $(f_n)_{n \in \mathbb{N}}$ be any $\ell_\mu^\infty$-computable sequence. If the sequence $(x_n)_{n \in \mathbb{N}} : x_n = \Psi(f_n)$ is well-defined (that is, if $\Psi(f_n) < \infty$ for all $n \in \mathbb{N}$), it is $\Sigma_2$-computable.*
2) *Let $(x_n)_{n \in \mathbb{N}}$ be any $\Sigma_2$-computable sequence and consider $\tau \in \mathbb{Q}$, $\tau > 1$. There exists an $\ell_\mu^\infty$-computable sequence $(f_n)_{n \in \mathbb{N}}$ s.t. for all $n \in \mathbb{N}$, we have $\Psi(f_n) = \min\{x_n, \tau\}$.* □

As follows from Theorem 4, $\Psi$ is $\Sigma_2$-*complete*, i.e., it is structurally exactly as complex as all other $\Pi_2$-problems. Analogous to the case of $\Omega$, this immediately implies the *uncomputability* of $\Psi$ as a function.

Since $\Psi$ is $\Sigma_2$-complete, it is "dual" to $\Omega$ in terms of the arithmetic hierarchy of real numbers. This duality is analogous to the duality of, respectively, NP and co-NP within the PSPACE complexity hierarchy. Note, however, that $\Omega$ and $\Psi$ are complete with respect to the *second* level of the arithmetic hierarchy of real numbers, while NP and co-NP form the *first* level of the PSPACE complexity hierarchy.

## VI. Conclusion

Provable integrity of digital-twinning systems has served as primary motivation for our framework. Technological systems – specifically those that involve physical processes in the sense of Section II – are commonly characterized through abstract real-valued mathematics. As indicated in Sections II and III, the computable-analysis approach provides formal guarantees concerning the behavior of such systems.

---

[3] For all $n \in \mathbb{N}$, we have $f_{n,*} \in \ell_\mu^\omega$. However, $(f_{n,*})_{n \in \mathbb{N}}$ is not *computable* (as a sequence) in $\ell_\mu^\omega$.

Modern programming languages for scientific computing often include functionalities that allow for the implementation of rudimentary methods from computable analysis. An example is the *BigInt* data type in *Julia*, based on which we can easily define computable numbers. To the best of the authors' knowledge, this is – beyond experimental and niche domains – the current state of affairs. Comprehensive implementations of computable-analysis tools have yet to reach the engineering practice. Formal methods, in contrast, are an active field in technological research and development. The relevant literature emphasizes their significance for safety-critical systems, where formal guarantees concerning systems behavior are obligatory (see e.g. [28]–[30]). However, the application domain of traditional formal methods consists of discrete state-based systems; extensions to real-valued physical problems often employ the RRAM setting. Upon implementation on digital hardware, such cases fall short of their rigorous mathematical validity (c.f. Section II).

From Section III, recall that computable analysis exhibits a *type-theoretic* formalization. At least to some extent, it is thus possible to mathematically unify computable analysis and common formal methods. While the underlying theory is well understood [10], establishing actual software implementations of such a unifying approach does not seem to have sparked widespread interest in the engineering community. Most relevantly, formal methods based on computable analysis would allow for analyzing and controlling physical processes without facing the RRAM setting's drawbacks.

Compared to the RRAM setting, the computable-analysis approach's arguably most immediate disadvantage is the significant increase in complexity, both mathematically and practically. Nevertheless, the authors believe that computable analysis ultimately forms the correct framework for analyzing, designing, and operating technological systems that require provable guarantees concerning their behavior.